# Extended infrared photoresponse in Te-hyperdoped Si at room temperature

Mao Wang[1,2,*], Y. Berencén[1], E. García-Hemme[3], S. Prucnal[1], R. Hübner[1], Ye Yuan[1,2], Chi Xu[1,2], L. Rebohle[1], R. Böttger[1], R. Heller[1], H. Schneider[1], W. Skorupa[1], M. Helm[1,2] and Shengqiang Zhou[1,*]

[1]Helmholtz-Zentrum Dresden-Rossendorf, Institute of Ion Beam Physics and Materials Research, Bautzner Landstr. 400, 01328 Dresden, Germany

[2]Technische Universität Dresden, 01062 Dresden, Germany

[3]Univ. Complutense de Madrid, Dpto. de Estructura de la Materia, Física Térmica y Electrónica 28040 Madrid, Spain

## Abstract

Presently, silicon photonics requires photodetectors that are sensitive in a broad infrared range, can operate at room temperature, and are suitable for integration with the existing Si technology process. Here, we demonstrate strong room-temperature sub-bandgap photoresponse of photodiodes based on Si hyperdoped with tellurium. The epitaxially recrystallized Te-hyperdoped Si layers are developed by ion implantation combined with pulsed laser melting and incorporate Te dopant concentrations several orders of magnitude above the solid solubility limit. With increasing Te concentration, the Te-hyperdoped layer changes from insulating to quasi-metallic behavior with a finite conductivity as the temperature tends to zero. The optical absorptance is found to increase monotonically with increasing Te concentration and extends well into the mid-infrared range. Temperature-dependent optoelectronic photoresponse unambiguously demonstrates that the extended infrared photoresponsivity from Te-hyperdoped Si *p-n* photodiodes is mediated by a Te-intermediate band within the upper half of the Si bandgap. This work contributes to pave the way towards establishing a Si-based broadband infrared photonic system operating at room temperature.

[*]Corresponding author, email: m.wang@hzdr.de; s.zhou@hzdr.de.

# I. INTRODUCTION

Presently, there is need for overcoming the bottleneck in the processing of the huge volume of data transmitted over the traditional telecommunication wavelengths around 1.3 µm and 1.5 µm [1,2]. Extending the spectral range from the near infrared (NIR) (from ~0.8 to ~2 µm) to the mid-infrared (MIR) (from ~2 to 25 µm) [3] has been proposed as a viable route for solving this setback [4-6]. Si-based photodetectors, compared with III-V (e.g. InSb) or HgCdTe infrared ones, satisfy the demand for cost-effective and environmental friendly solutions, and enable the development of on-chip complementary-metal-oxide-semiconductor (CMOS)-compatible photonic systems [5,7,8]. However, their photoresponse is fundamentally limited in the visible and near infrared spectral regime owing to the relatively large band gap of Si (1.12 eV, λ= 1.1 µm). Therefore, the room-temperature broadband infrared detection is of great interest in the realm of silicon photonics. Particularly, hyperdoped Si materials are on the spotlight of present-day investigations due to their superior optoelectronic properties (e.g. the highest absorption coefficient (~$10^4$ cm$^{-1}$) ever obtained for Si in the infrared range, which is comparable to that of intrinsic Ge) after incorporating certain concentration and type of dopants [9]. More recently, hyperdoping has successfully been extended to nanostructured Si such as nanowires and nanocrystals [10,11] where localized surface plasmon resonances [12] and sub-bandgap optoelectronic photoresponse have been demonstrated [13]. This new type of material also excels Si/Ge [14] and Si/III-V semiconductor heterostructures [15] from the standpoint of fabrication [16]. However, extrinsic IR photodetectors based on Si with dopants of group III (B, Al and Ga) and V (P, As and Sb) can only operate at temperatures below 40 K [7,17-19], since the introduced shallow-dopant levels within the Si band-gap are thermally ionized at room temperature.

Alternatively, Si photodetectors that make use of deep level dopants (with a much higher thermal ionization energy), such as Ti, Ag, Au, S and Se [20-25], are effective for strong room-temperature photoresponse, extending well in the infrared range [23,24,26,27]. The extended photoresponse has been demonstrated to be associated with known dopant deep-energy levels within the Si band-gap. However, less attention has been paid to the optoelectronic photoresponse of Te-hyperdoped Si [28]. In fact, chalcogens are double donors in Si, and especially Te offers important advantages for hyperdoping Si over its chalcogen counterparts, namely S and Se dopants. For instance, Te introduces deep donor states at $E_c$- 0.20 eV and $E_c$-0.41 eV, respectively, which are located in the upper half of the Si band-gap and in turn are closer to the conduction band (CB) compared with S and Se as schematically shown in FIG. 1 [29,30]. This feature of Te facilitates to further extend the photoresponse towards the MIR range. However, Te has a larger atomic radius and is much less electronegative than S and Se (S-2.58, Se-2.55, Te-2.10) [31]. The latter results in a larger substitutional formation energy ($E^f$) of Si:Te, e.g. $E^f[Si_{125}:S]$=-0.13 eV, $E^f[Si_{125}:Se]$=+1.29 eV and $E^f[Si_{125}:Te]$=+1.97 eV [32]. Thus, achieving a hyperdoping level in Si:Te comparable to the one reported for Si:S or Si:Se is more challenging.

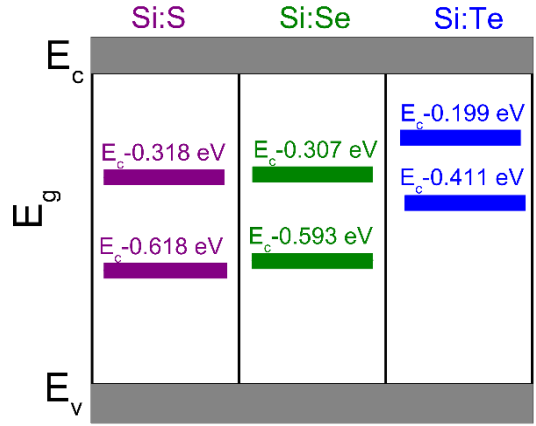

FIG. 1. Energy-level schemes for S-, Se-, and Te-doped in Si [29,30,32,33]. (Color online)

In this work, we report on the realization of single-crystalline and epitaxial hyperdoped Si with Te concentrations as high as $10^{21}$ cm$^{-3}$ by ion implantation combined with pulsed laser melting (PLM). The observed quasi-metallic behavior is driven by increasing Te concentration. Te-hyperdoped Si *p-n* photodiodes exhibit a remarkably broad spectral range down to 3.1 μm at room temperature with an enhanced optoelectronic photoresponse compared to the Au-hyperdoped Si-based photodetectors [23]. This work points out the potential of Si hyperdoped with Te for room-temperature infrared detection as the new generation of Si-based photonic systems.

## II. EXPERIMENTAL DETAILS

Double-side polished Si (100) wafers (*p*-type, boron-doped, $\rho \approx$ 1-10 Ωcm) were implanted with Te ions at six different fluences at room temperature (see table 1). The Te depth profile was firstly calculated using SRIM code [34] and then verified by Rutherford backscattering spectrometry/channeling (RBS/C) measurements. A combined implantation with energies of 150 keV and 50 keV with a fluence ratio of 2.5:1 was applied to obtain a uniform distribution of Te in the implanted layer. Subsequently, ion-implanted samples were molten using a pulsed XeCl excimer laser (Coherent COMPexPRO201, wavelength of 308 nm, pulse duration of 28 ns) in ambient air. A single laser pulse of beam size 5 mm × 5 mm with the fluence of 1.2 J/cm$^2$ was used to anneal the as-implanted layers. The choice of this PLM-fluence was carried out by inspecting the crystalline quality and the Te depth profile of the layers using Raman and RBS measurements, respectively. During the annealing process, the whole amorphous implanted region was molten and then recrystallized with a solidification speed in the order of 10 m/s while cooling down [35]. This allows for Te concentrations beyond the solid solubility limit of Te in Si while preserving the epitaxial single-crystal growth.

TABLE I. Sample description. The depth distribution of Te (estimated thickness=120 nm) is calculated using SRIM and verified by RBS measurements (A fit to the RBS random spectrum using the SIMNRA code [36,37] yields the Te concentration.) The sample names refer to the peak Te concentration of the as-implanted layer.

| Sample ID | Implantation parameters | Tellurium peak concentration (%) |
|---|---|---|
| Te-0.25% | 150 keV, 7.8×10$^{14}$ cm$^{-2}$; 50 keV, 3.1×10$^{14}$ cm$^{-2}$ | 0.25 |
| Te-0.50% | 150 keV, 1.6×10$^{15}$ cm$^{-2}$; 50 keV, 6.2×10$^{14}$ cm$^{-2}$ | 0.50 |
| Te-1.0% | 150 keV, 3.1×10$^{15}$ cm$^{-2}$; 50 keV, 1.2×10$^{15}$ cm$^{-2}$ | 1.0 |
| Te-1.5% | 150 keV, 4.7×10$^{15}$ cm$^{-2}$; 50 keV, 1.9×10$^{15}$ cm$^{-2}$ | 1.5 |
| Te-2.0% | 150 keV, 6.2×10$^{15}$ cm$^{-2}$; 50 keV, 2.5×10$^{15}$ cm$^{-2}$ | 2.0 |
| Te-2.5% | 150 keV, 7.8×10$^{15}$ cm$^{-2}$; 50 keV, 3.1×10$^{15}$ cm$^{-2}$ | 2.5 |

To analyze the microstructure of the PLM-treated Te-hyperdoped Si layers, high-resolution TEM (HRTEM) was performed on an image $C_s$-corrected Titan 80-300 microscope (FEI) operated at an accelerating voltage of 300 kV. High-angle annular dark-field scanning transmission electron microscopy (HAADF-STEM) imaging and spectrum imaging based on energy-dispersive X-ray spectroscopy (EDXS) were done at 200 kV with a Talos F200X microscope equipped with an X-FEG electron source and a Super-X EDXS detector system (FEI). Prior to STEM analysis, the specimen mounted in a high-visibility low-background holder was placed for 10 s into a Model 1020 Plasma Cleaner (Fischione) to remove contamination. Classical cross-sectional TEM lamella preparation was done by sawing, grinding, polishing, dimpling, and final Ar ion milling. RBS/C measurements were performed with a collimated 1.7 MeV He$^+$ beam of the Rossendorf van de Graff accelerator with a 10-20 nA beam current at a backscattering angle of 170° to investigate the crystalline quality of the pulsed-laser annealed Te-hyperdoped Si layers. The channeling spectra were collected by aligning the sample to make the impinging He$^+$ beam parallel to the Si [001] axes.

The conductivity type, carrier concentration and carrier mobility of the PLM-treated Te-hyperdoped layers were measured by a commercial Lake Shore Hall measurement system in a van der Pauw configuration [38] under a magnetic field perpendicular to the sample plane. The magnetic field was swept from -4 T to 4 T. The gold electrodes were sputtered onto the four corners of the square-like samples. The native SiO$_2$ layer was removed by HF etching prior to the sputtering process. Next, a silver conductive glue paste was used to contact the wires to the gold electrodes. All contacts were confirmed to be ohmic by measuring the current-voltage curves at different temperatures.

Transmittance ($T$) and reflectance ($R$) measurements were performed at room temperature by means of a Fourier transform infrared spectroscopy (FTIR) using a Bruker Vertex 80v FT-IR spectrometer to quantify the sub-bandgap absorptance of the Te-hyperdoped Si layers in the infrared spectral range of 0.05-0.85 eV ($\lambda$ = 1.4-25 μm). To this end, the probe beam was focused on the PLM-treated area of the sample (5 mm × 5 mm). The absorptance ($A = 1-T-R$) was then determined by recording the transmittance and reflectance spectra.

The temperature-dependent photoresponsivity from 60 K to 300 K was measured in the Te-hyperdoped Si/$p$-Si photodiode devices. A 0.068 cm$^2$ illuminated area was photolithographically processed by defining fingers with a separation of 200 μm resulting in frame-like Ti/Al top electrodes on top of the Te-hyperdoped Si layer. The bottom-side electrode was made by Ti/Al contact on the nearly entire bottom surface, but avoiding sample edges in order to reduce possible parasitic electrical conduction through the edges of the sample. Both the top-side and bottom-side electrodes were defined by e-beam evaporation using a 50 nm/100 nm Ti/Al bilayer. The photoresponsivity measurements were conducted by placing the $p$-$n$ photodiodes inside a helium closed-cycle Janis cryostat with a ZnSe window. A vacuum pump was used to avoid moisture condensation at low temperatures [22]. A SiC glowbar infrared source coupled to a TMc300 Bentham monochromator equipped with gratings in Czerny-Turner reflection configuration were used. The incident 1.5 cm × 3 cm light spot, spatially homogeneous in intensity, was used to study the spectral photoresponse of the device. The system was calibrated with a Bentham pyrometric detector. The short-circuit photocurrent between the top and botton contacts was measured with a SR830 DSP lock-in amplifier. For all the measurements, the infrared light from the SiC source was mechanically chopped at 87 Hz before entering the monochromator.

## III. EXPERIMENTAL RESULTS
### A. Structural Properties

A representative cross-sectional HAADF-STEM micrograph superimposed with the corresponding Te, Si, and O element maps is shown in FIG. 2(a) for the PLM-treated Te-hyperdoped Si layer (Te-1.5%). Apart from the native oxide layer at the surface, Te is found to be evenly distributed within the top 125 nm of the Si wafer. Neither Te surface segregation nor nano-scale Te agglomerates are detected despite the high doping concentration of more than $10^{20}$ cm$^{-3}$. Single-crystalline regrowth of the Te-hyperdoped Si layer during the PLM treatment is confirmed by HRTEM imaging and the subsequent Fast Fourier transform analysis as shown in FIG. 2(b). Moreover, extended defects, secondary phases or cellular breakdown are not observed.

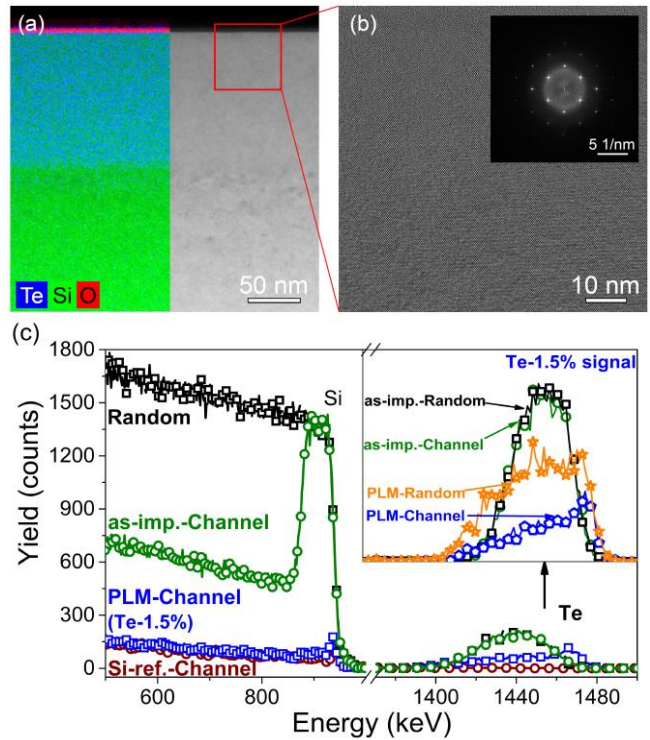

FIG. 2. Microstructure investigation of Te-hyperdoped Si layers (color online). (a) Cross-sectional HAADF-STEM image superimposed with the corresponding EDXS element maps (blue: tellurium, green: silicon, red: oxygen) for the PLM-treated Te-hyperdoped Si layer Te-1.5%; (b) representative HRTEM image with corresponding Fast Fourier transform (inset) for a field of view as depicted by the red square in image part (a); (c) a sequence of 1.7 MeV He RBS/C spectra of PLM-treated Te-hyperdoped Si layer. The inset in FIG. 2 (c) shows the magnification of random (-R) and channeling (-C) Te signals from as-implanted and PLM-performed Te-hyperdoped Si samples with a concentration of 1.5%. (Color online)

After the PLM treatment, all the Te-hyperdoped Si samples show a similar crystalline quality and Te depth distribution. FIG. 2(c) depicts the representative RBS/C spectra of a PLM-treated Te-hyperdoped Si layer together with the virgin Si and the as-implanted sample. The random spectrum of the as-implanted layer reveals a thickness of about 120 nm for the Te profile, in an approximately Gaussian distribution. No channeling effect is observed in the implanted layer because of the amorphization caused by ion implantation. From the RBS-channeling signal, the PLM-treated layer shows a minimum backscattered yield $\chi_{min}$ of 4% (defined as the ratio of the aligned to the random yields), which is comparable to that of the virgin Si substrate. This indicates a full re-crystallization and an epitaxial growth of the PLM-treated layer, in agreement with the HRTEM results in FIG. 2(b).

It is evidenced in the Te signal that Te atoms tend to diffuse towards both the surface and the substrate sides during the PLM process, leading to a relatively uniform Te profile with a thickness of 125 nm, as illustrated in FIG. 2(a). Moreover, the inset in FIG. 2(c) shows that the channeling spectrum of the 1.5% Te hyperdoped sample exhibits a $\chi_{min}$ of about 4% for Si and 30% for Te in the Te-implanted region. The substitutional fraction for Te (i.e., the ratio of substitutional Te dopants at the Si lattice sites to the total implanted Te atoms) can be estimated as (1–30%)/(1–4%)=72% [39]. The substitutional fraction of all the Te-hyperdoped Si samples is around 70%, higher than that of Se in Si [40].

### B. Electrical characterization

Electrical measurements were made to investigate the transport properties of the PLM-treated Te-hyperdoped Si layers. To avoid the influence of the parallel conduction from the *p*-type Si substrate, a set of samples with the same range of Te concentrations were processed on an *intrinsic* Si substrate ($\rho > 10^4$ Ωcm). FIG. 3 shows the temperature-dependent resistivity of Te-hyperdoped Si samples with different Te concentrations. At room temperature, the resistivity of the Te-hyperdoped Si layers is less than $10^{-2}$ Ωcm, which is much lower than that of the Si substrate. This confirms that the intrinsic Si substrate has no influence on the transport properties of the Te-hyperdoped Si layer considering the respective thickness. FIG. 3 shows that the resistivity decreases with increasing Te concentration and a remarkable difference in a factor of $10^4$ is reached at 2 K. Samples with 0.25% and 0.50% of Te content behave like an insulator with the resistivity sharply increasing at low temperatures. This is because of electrons return to their localized ground states from the thermally excited conduction-band states as temperature decreases. However, samples with Te concentrations higher than 1% exhibit a different behavior as shown in the inset of FIG. 3, i.e. their resistivity slightly increases as the temperature decreases. Samples with a higher doping concentration show quasi-metallic behavior, i.e. their conductivity remains finite as the temperature tends to zero due to the delocalization of donor electrons above a critical donor concentration ($n_c$) [41]. This gradual transition from insulating to quasi-metallic behavior driven by the Te concentration involves the formation of a broad intermediate band (IB) in the upper half of the Si bandgap [42,43].

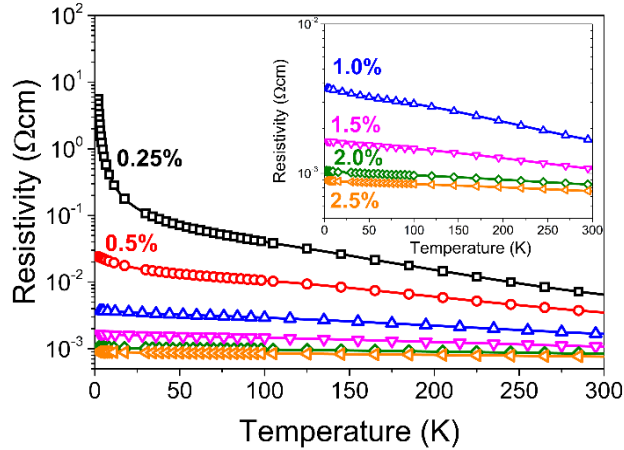

FIG. 3. Temperature-dependent resistivity of the Te-hyperdoped Si layers with different Te concentrations. The inset shows a magnification of the data for samples with higher Te concentration. (Color online)

## C. Optical absorption

A strong and broad sub-bandgap optical absorptance in Te-hyperdoped Si layers, as compared with a bare Si sample, is shown in FIG. 4. The sharp peaks below 0.15 eV in the absorptance spectra are known to be related with oxygen and carbon impurities [44,45] in the Si substrate. The absorptance extends down to 0.048 eV (the MIR detection limit of FTIR) while it increases with the Te concentration, which is consistent with the previously reported results about S- and Se- hyperdoped Si [27,46]. The well-defined broad absorptance band peaking at around 0.36 eV for samples Te-0.25% and Te-0.5% comes from the presence of discrete impurity levels or a narrow intermediate band. The peak position correlates well with the activation energy of the deep Te-levels [29] and agrees with previous works [32,47]. On the other hand, samples with Te concentrations in excess of 1% show quasi-metallic behavior and the broad absorptance band evolves leading to a strong increase of the absorptance down to 0.048 eV. This can be understood in terms of a free-carrier absorption process because of the high carrier concentration in the Te-hyperdoped Si layers ($\geq 10^{20}$ cm$^{-3}$). For the sample with the highest Te content, the decrease of the sub-bandgap absorptance is likely due to the formation of inactive Te-$Vm$ (vacancy-impurity complexes) with similar complexes as reported in As-hyperdoped Si [48] since the Si host cannot any longer accommodate more Te atoms at such high concentrations ($1 \times 10^{21}$ cm$^{-3}$).

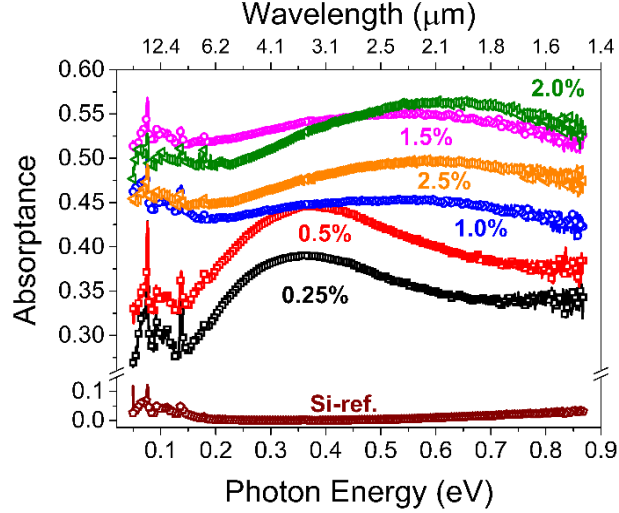

FIG. 4. Room-temperature sub-bandgap optical absorptance spectra from PLM-treated Te-hyperdoped Si samples with different Te concentrations. A bare Si substrate is shown as a reference. (Color online)

### D. Infrared optoelectronic response

The sub-bandgap photoresponse in Te-hyperdoped Si layers is further investigated via the spectral responsivity (A·W$^{-1}$) of $p$-$n$ photodiode devices. The top view and cross-section of the Te-hyperdoped Si/$p$-Si photodiode are schematically presented in FIG. 5(a) and (b), respectively. FIG. 5(c) shows an I-V curve at room-temperature in dark, FIG. 5(d) depicts the infrared responsivity (A·W$^{-1}$), which is estimated at zero bias (i.e. photovoltaic mode), as a function of incident photon energy. Measurements are performed at zero bias in order to prove the pure photovoltaic effect of the Te-hyperdoped Si/$p$-Si photodiodes. This will rule out any other possible origin of the photoresponse. Measurements under a reverse bias could give a false positive such as a photoconductive effect. The zero-bias room-temperature responsivity of a commercial Si PIN photodiode (model: BPW34) is also included for comparative purposes in FIG. 5(d). We fabricated the photodiodes on the Te-hyperdoped Si sample with the strongest sub-bandgap absorptance (see FIG. 4, sample Te-2.0%).

Figure 5(c) shows that the fabricated photodiode presents a rectifying behaviour. The forward voltage corresponds to positive bias applied to the bottom contact ($p$-Si substrate) and the ratio of direct/reverse current is 36 at ±0.5 V. To further analyze the rectifying behaviour, the dark IV curve was fitted to a single-diode model [49] using:

$$I = I_0 \left[ e^{\frac{q(V-IR_S)}{\eta k_B T}} - 1 \right] + \frac{V-IR_S}{R_{shunt}} \quad (1),$$

where $I_0$ is the saturation current, $q$ is the electron charge, $T$ is the temperature, $\eta$ is the ideality factor and $k_B$ is the Boltzmann constant. $V$ and $I$ are the total voltage and current whereas $R_s$ and $R_{shunt}$ are the series and parallel resistances, respectively. Hence, $R_s$, $R_{shunt}$, $I_0$ and $\eta$ were found to be 6.4 Ω, 850 Ω, 68 µA and 2.2, respectively. The deduced series resistance value is in agreement with the calculated series resistance, $R_s = \rho \frac{t}{A}$, assuming a wafer resistivity of $\rho$=1-10 Ωcm, a wafer thickness of $t$=380 µm and an electrode area of A=0.068 cm². This confirms the ohmic character of the contacts. On the other hand, an ideality factor of 2.2 suggests i) the existence of a *p-n* junction rather than a metal-semiconductor junction, in which the ideality factor is expected to be 1 [50] and ii) that the main conduction mechanism is ascribed to a recombination/generation process of carriers in the depletion region of the *p-n* photodiode [49]. Therefore, the observed rectifying behaviour in FIG. 5(c) is directly related to the *p-n* junction between the *p*-type Si substrate and the *n*-type Te-hyperdoped Si layer.

FIG. 5(d) depicts the temperature-dependent responsivity as a function of photon energy. The Te-hyperdoped Si *p-n* photodiode shows a strong sub-bandgap responsivity down to 0.3 eV in the whole temperature range. At room-temperature, the Te-hyperdoped Si *p-n* photodiode exhibits a responsivity of around $10^{-4}$ AW$^{-1}$ at the two telecommunication optical wavelengths (1.3 µm and 1.5 µm), whereas the responsivity at photon energies below 0.9 eV of a commercial Si PIN photodiode reaches the noise floor as expected. The measured responsivity of $10^{-4}$ AW$^{-1}$ is comparable to the ones reported for hyperdoped Si-based photodiodes [23] and solar cells [51] at the corresponding wavelengths. Moreover, the external quantum efficiency (EQE) at 1.3 µm and 1.5 µm is $2\times10^{-4}$ and $8\times10^{-5}$, respectively, as estimated by:

$$\text{EQE}(\lambda) = \frac{R_{ph}}{\lambda} \times \frac{hc}{e} \approx \frac{R_\lambda}{\lambda} \times (1240\ W \cdot nm\ /A) \quad (2),$$

where $R_{ph}$ is the spectral responsivity (i.e. the ratio of the electrical output to the optical input.), $\lambda$ is the wavelength in nm, $h$ is the Planck constant, $c$ is the speed of light in vacuum, and *e* is the elementary charge. The estimated EQE values are comparable to other deep level impurity-hyperdoped Si photodiodes [23,27].

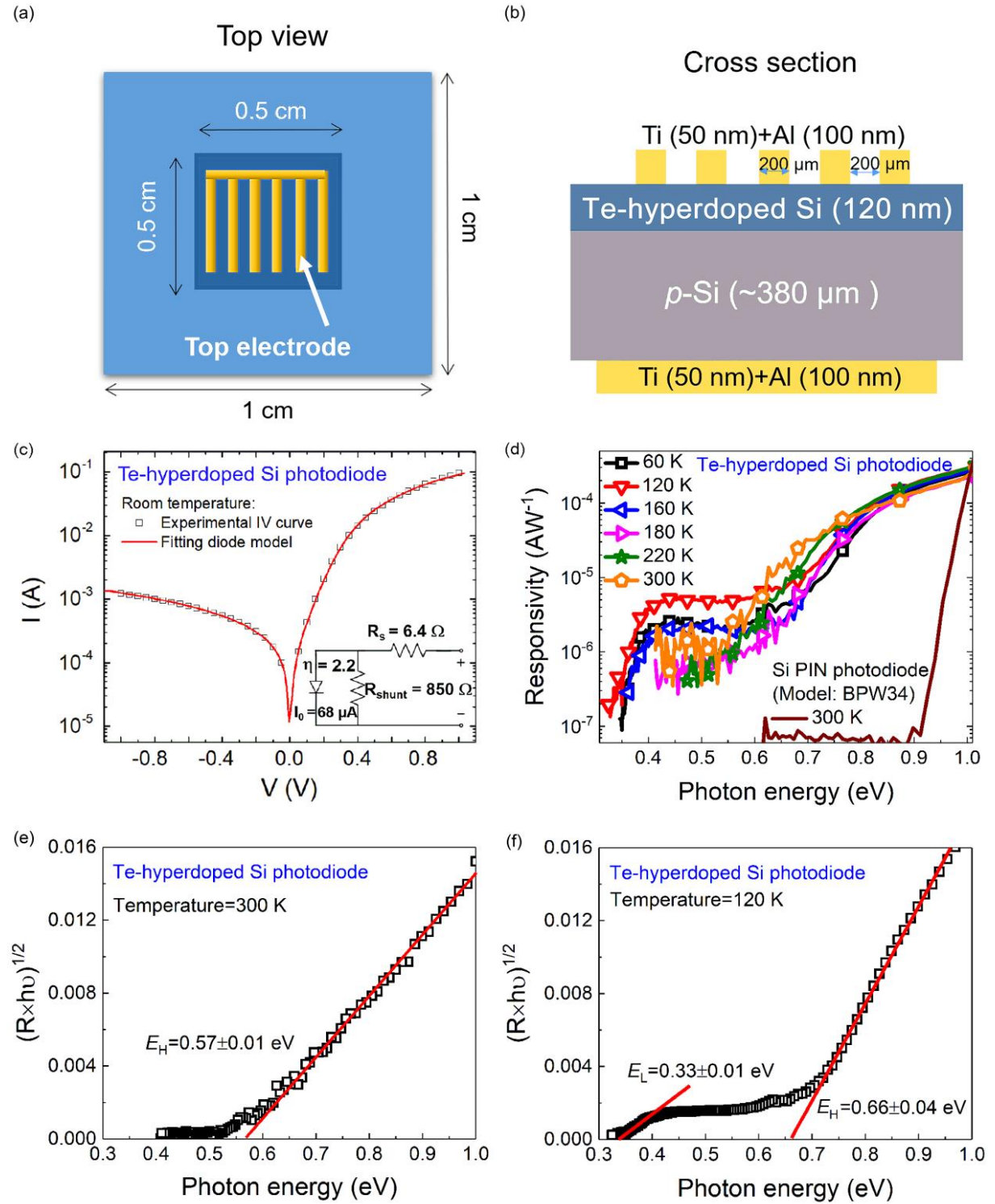

FIG. 5 (a) Top view and (b) cross-section scheme of the Te-2.0%-hyperdoped Si/*p*-Si photodiode devices. (c) The dark current as a function of applied bias voltage measured at room temperature. The inset in FIG. 5(c) shows the equivalent electrical circuit of the experimental setup. (d) The spectral responsivity measured at zero bias (i.e. photovoltaic mode) for the Te-hyperdoped Si photodiode and the commercial Si PIN photodiode at different temperatures. (e) and (f) show the fitting of the absorption edge using Eq. (3) at different temperature ranges. (Color online)

Regarding the responsivity at different temperatures of the Te-hyperdoped Si *p-n* photodiode, two ranges of temperatures with different behaviours in terms of the line shape and responsivity intensity can be identified. At high-temperature range (300 K to 180 K), the responsivity extends down to ~0.55 eV, where the noise-floor is reached. However, at low-temperature range (160 to 60 K), an additional broad photoresponse band spanning from 0.60 eV to 0.33 eV is clearly observed. These features allow us to define a photoresponse band at high photon energies $E_H$, that lies in the range of 0.55-0.65 eV and another one at low photon energies $E_L$, in the range of 0.3-0.4 eV.

To get insight on the optical transitions coming from the Te-related intermediate band, a direct proportionality between the photoresponsivity and the absorption coefficient at energies close to the optical transition edges was assumed. The sub-bandgap optical transitions can therefore be derived by the Tauc method using the following modified expression:

$$R_{ph} \times h\nu = A(h\nu - E_g)^n \qquad (3)$$

where $h\nu$ is the photon energy, $A$ is a constant, $E_g$ stands for the Si bandgap and $n$ would adopt the value of 2 for the indirect optical transitions. A Tauc-type plot for both the high-temperature and the low-temperature ranges at two representative temperatures (viz. 300 K and 120 K) is shown in FIG. 5(e) and (f), respectively. The energy gap derived from this plot is determined to be $E_H$=0.57±0.01 eV at 300 K and $E_H$=0.66±0.04 eV at 120 K, respectively. Alternatively, a second absorption edge band is found to be $E_L$=0.33±0.01 eV at 120 K. The slight difference between the deduced $E_H$ values at 300 K and 120 K are thought to be related with the temperature dependence of the Si bandgap.

Next, we recorded the device photocurrent as a function of the input optical power at photon energies of 0.82 eV and 0.42 eV at 300 K and 120 K, respectively (FIG. 6). The photocurrent in response to the sub-bandgap illumination at 300 K and 120 K is found to scale linearly with the input optical power in a log-log representation with a slope of around 1. This suggests that the sub-bandgap photoresponse mechanism of the Te-hyperdoped Si *p-n* photodiodes is related to a single-photon absorption process mediated by the intermediate band rather than a two-photon absorption process. In addition, the possibility of internal photoemission from the top electrode is discarded by the proven ohmic character of the contacts.

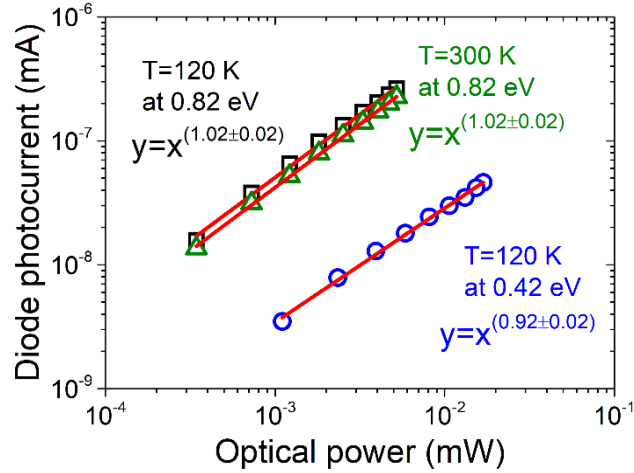

FIG. 6. Diode photocurrent versus input optical power with the corresponding linear fits at the photon energies of 0.82 eV and 0.42 eV at 120 K, and at 0.82 eV at 300 K, respectively. (Color online)

## IV. DISCUSSION

We discuss on the origin of the sub-bandgap photoresponsivity at room- and low-temperatures from the Te-hyperdoped Si *p-n* photodiodes. The results derived from the analysis of the material properties in conjunction with the spectral photoresponsivity at different temperatures are consistent with the formation of a Te-related intermediate band in the upper half of the Si bandgap. The Te intermediate band facilitates the generation of charge carriers originated by the absorption process of photons with energies lower than the Si bandgap. For the sake of clarity, FIG. 7 shows a sketch of the band diagram of the Te intermediate band-based Si *p-n* photodiode with the different involved processes in the sub-bandgap optoelectronic photoreponse:

i) In the high-temperature range (300 K to 180 K), the transition from the IB to the CB (process I), denoted as $E_L = E_C$-0.33 eV, takes place but the ratio of thermal to optical carrier generation is sufficiently high to screen a measurable photocurrent arising from this transition. In detail, an equilibrium conduction-band electron concentration coming from the thermal carrier generation as high as $10^{17}$ cm$^{-3}$ can be estimated by the charge carrier statistics [49]. This, however, does not impede the occurrence of the transition from the VB to the IB (Process II), which is in accordance with the kink in the room-temperature spectral photoresponsivity at $E_H = 0.57\pm0.01$ eV as shown in FIG. 5(e).

ii) In the low-temperature range (160 K to 60 K), the thermal carrier generation is then suppressed by the freeze-out effect [52]. This leads to a ratio of thermal to optical carrier generation low enough that allows for a measurable photoresponsivity band arising from the IB to the CB (process I) at $E_L = 0.33\pm0.01$ eV (FIG. 5(f)). Likewise, the transition from the IB to the CB (process I) gives rise to available states at the IB which are then populated by electrons from the valence band. This results in the photoresponsivity band at $E_H = 0.66\pm0.04$ eV (FIG. 5(f)). The observation of the two sub-bandgap optical transitions (VB→IB and IB→CB) indicates that the IB is not merged with the CB despite the high Te concentration.

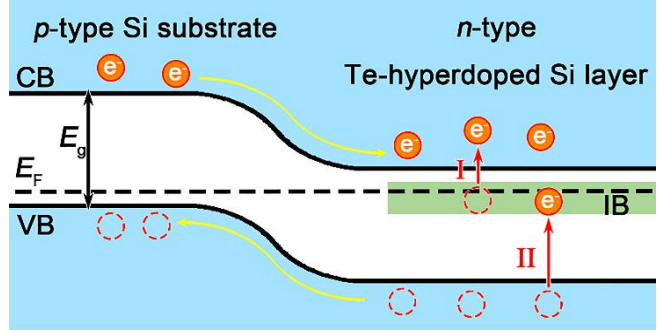

FIG. 7 Band diagram of the Te intermediate band-based Si *p-n* photodiode. Te dopants introduce electron states (intermediate band) of about 0.33 eV below the conduction band of Si. The Te intermediate band facilitates the generation of charge carriers that participate in the absorption of two or more sub-bandgap-energy photons. The transition of IB to CB (process I) takes place at $E \geq E_L$ while is only measureable at low temperature because the almost no contribution from the thermal generation process. The transition of VB to IB (process II) occurs at $E \geq E_H$. (Color online)

While the high-quality hyperdoped Si material platform is established, further efforts must be destined towards an advanced device design to boost the device efficiency of this prototype of photodiode. This can be achieved by designing a top electrode with narrower fingers and gaps that helps to enhance the carrier collection efficiency. An antireflection coating layer in conjunction with a passivation layer might also help to improve the efficiency.

## V. CONCLUSION

The possibility of generating photocarriers at room temperature using sub-bandgap radiation in high-quality single-crystal Si films doped with Te concentrations greater than the thermal solubility limit has been established. The Te-hyperdoped Si films exhibit a broad optical absorptance spanning from 1 μm to 13 μm. A clear evolution from the discrete energy levels to the formation of an intermediate band as a function of Te concentration has been demonstrated. The extended infrared photoresponse of the Te-hyperdoped Si *p-n* photodiodes has been proven to be mediated by an intermediate band within the upper half of the Si bandgap. The CMOS-compatible approach demonstrated here provides a way to achieve room-temperature sub-bandgap optoelectronic response in single-crystal Si-based intermediate band materials.

# ACKNOWLEDGMENTS

Authors acknowledge the ion implantation group at HZDR for performing the Te implantations. Additionally, support by the Structural Characterization Facilities at Ion Beam Center (IBC) and funding of TEM Talos by the German Federal Ministry of Education of Research (BMBF), Grant No. 03SF0451 in the framework of HEMCP are gratefully acknowledged. This work is funded by the Helmholtz-Gemeinschaft Deutscher Forschungszentren (HGF-VH-NG-713). M.W. thanks financial support by Chinese Scholarship Council (File No. 201506240060). Y.B. would like to thank the Alexander-von-Humboldt foundation for providing a postdoctoral fellowship. E.G.H would like to thank the Spanish MINECO (Ministerio de Economía y Competitividad) for the financial support under grant TEC2017-84378-R.